      [1999/12/01 v1.4c Il Nuovo Cimento]
\documentclass{cimento}

\usepackage{subfigure}
\usepackage{graphicx}  % got figures? uncomment this
\title{\texttt{MCSTHAR++}, a Monte Carlo code for the microcanonical hadronization}
\author{C.~Bignamini\from{ins:x}\thanks{Speaker.\\ 
Presented at IFAE 2010, Roma, Italy, 7-9 Apr 2010.\\ \\
Work supported in part by the the EU Marie Curie Research Training Network
MCnet (MRTN-CT-2006-035606).
}\ETC,
F.~Becattini\from{ins:y} \atque
F.~Piccinini\from{ins:z}
}
\instlist{	\inst{ins:x} Dipartimento di Fisica Nucleare e Teorica, Universit\`a di Pavia, and INFN Sezione di Pavia - Pavia, Italy
	\inst{ins:y} Dipartimento di Fisica, Universit\`a di Firenze, and INFN Sezione di Firenze - Firenze, Italy
	  \inst{ins:z} INFN Sezione di Pavia - Pavia, Italy}
\PACSes{
\PACSit{12.40.Ee}{Statistical models of strong interactions}
\PACSit{13.66.Bc}{Hadrons production by electron-positron collisions}}
\begin{document}

\maketitle

\begin{abstract}
%\texttt{MCSTHAR++} is a new Monte Carlo code performing the hadronization process as described by the Statistical Hadronization Model 
%in the microcanonical framework, whose main feature is the minimal number of free parameters. \\ 
%\texttt{MCSTHAR++} can be used in the Monte Carlo event generators as an alternative to the standard hadronization modules. \\    
%The study of the capabilities of this model to reproduce experimental distributions is ongoing.
\texttt{MCSTHAR++} is a new Monte Carlo code implementing the
Statistical Hadronization Model. This model assumes that hadronization
proceeds through the microcanonical decay of massive extended clusters.
Unlike other hadronization models, in this approach very few free
parameters are needed, as has been demonstrated in previous
studies. The tuning of the model and the comparison with the data is
ongoing.
\end{abstract}

\section{Introduction} 
The transition from the partonic state to the confined state is
described using various phenomenological models like the \emph{Cluster Models} \cite{ref:CluMod}\cite{ref:SherpMod} and the \emph{String Model} \cite{ref:StrMod}.
A completely independent model, based on a statistical formulation and known as \emph{Statistical Hadronization Model} \cite{ref:HagBec}, has been studied in the literature but it is not available in any official release of the Monte Carlo event generators for High Energy Physics. 
In this paper a short overview of the microcanonical formulation of the Statistical Hadronization Model will be given and a new Monte Carlo code
performing the hadronization process according to that model, \texttt{MCSTHAR++}, and its interface
to the event generator \texttt{HERWIG6.510} \cite{ref:Herwig65} will be described, showing also some preliminary results obtained for the hadronization
of light quarks only.

\section{The Statistical Microcanonical Hadronization Model: a short overview}
As described in \cite{ref:HagBec}, the statistical model is based on the hypothesis that in a high energy collision 
some extended objects made of pre-hadronic matter, called \emph{clusters} or \emph{fireballs}, are produced. These objects are
supposed to be colorless and characterized by well defined physical quantities like energy, momentum and abelian charges.
Each one of these clusters hadronizes according to a pure statistical law, which for the microcanonical formulation
is the following: \emph{every multihadronic state confined inside the cluster and conserving all the physical quantities of the cluster itself is equally likely}.
In this picture of the hadronization process the cluster is treated as a microcanonical ensemble, whose partition function $\Omega$ is given by \cite{ref:BecFer}

\begin{equation}
 \label{part0}
\Omega = \displaystyle\sum_{ \left(N_{j}\right)}\langle \left(N_{j}\right) \mid P_{\mathbf{Q}}P_{V}P_{\mathbf{Q}} \mid  \left(N_{j}\right)\rangle, \quad
                     \tx{where\ } P_{V} = \displaystyle\sum_{h_{V}}\mid h_{V}\rangle\langle h_{V}\mid
 \end{equation}

is the sum over all the multihadronic states $\mid h_{V}\rangle$ confined inside the cluster, while the sum on the left equation is over all possible channels, 
corresponding to the asymptotic states, which are identified by the $K$-tuple of integer
$\left(N_{j}\right) = (N_{1},N_{2},...,N_{K})$, one for each hadron species $j$, and where $P_{\mathbf{Q}}$ is the projector on the conserved set of charges. 
%This leads to the following explicit equation for the partition function

%\begin{equation}
% \label{eq:part1}
%\Omega = \displaystyle\sum_{\left(N_{j}\right)}{\frac{V^{N}}{\left(2\pi\right)^{3N}}\left(\prod_{i=1}^{N}}{\left(2J_{i}+1\right)\int{d^{3}p_{i}}}\right)\delta^{4}{\left(\mathbf{P}-\mathbf{P}_{\left(N_{j}\right)}\right)\delta_{\mathbf{Q},\mathbf{Q}_{\left(N_{j}\right)}}}},
%\end{equation}

%where $V$ is the volume of the cluster, $N$ is the number of particles of the channel $\left(N_{j}\right)$, $J_{i}$ the spin of each one of them and $\mathbf{P} \left(\mathbf{Q}\right)$ and $\mathbf{P}_{\left(N_{j}\right)} \left(\mathbf{Q}_{\left(N_{j}\right)\right)$ are the 4-momentum (charge configuration) of the cluster and of the channel $\left(N_{j}\right)$ respectively. The previous equation is the simplest form of the partition function, since 
%the Bose-Einstein and Fermi-Dirac correlations, as well as the interactions among the hadrons, are not included: this is the approximation used to obtain the following preliminary results, the complete formula, including also such effects, can be found in \cite{ref:BecFer}. 
The above partition function is then modified introducing the strangeness suppression parameter $\gamma_{s}$ and multiplying the weight of each channel by the factor $\gamma_{s}^{N_{s}}$, where $N_{s}$ is the total number of strange and antistrange quarks contained in the hadrons of the channel itself. At this point this hadronization model requires only \emph{two} parameters, which need to be tuned on the experimental data: the $\gamma_{s}$ parameter and the energy density of the clusters $\rho$, which is taken to be equal for each cluster and which is used to convert the mass of a cluster into its volume, a quantity which appears in $\Omega$, as can be seen in \cite{ref:BecFer}.
% where it is also described how the quantum correlations and the interactions, not included in the following results, can be taken into account.

\section{\texttt{MCSTHAR++}: code description and preliminary results}
\texttt{MCSTHAR++} (\textbf{M}onte \textbf{C}arlo \textbf{ST}atistical \textbf{HA}dronization in high energy \textbf{R}eactions) is a Monte Carlo code implementing the Statistical Hadronization Model in its microcanonical formulation. It is written in Object Oriented \texttt{C++} and it is built to take as input a set of clusters and to give  in output a set of hadrons (stable and unstable), coming from the microcanonical hadronization of each one of the incoming clusters. 
For the hadronization of these objects the exact conservation of energy-momentum, electric charge, strangeness, baryonic number, charm and beauty charge is imposed. 

In the present case, \texttt{MCSTHAR++} is interfaced to \texttt{HERWIG6.510} \cite{ref:Herwig65}, the clusters used as input  for \texttt{MCSTHAR++} are the "standard" \texttt{HERWIG}'s clusters (allowed to have non zero baryonic number) and the primary hadrons produced during the hadronization are decayed using the \texttt{HERWIG}'s routines for the strong and electroweak hadron decay.

Even if a full tuning of the model is mandatory, it is already possible to see the good behavior of the hadronization code in the preliminary results shown in this section. The theoretical predictions are obtained with no tuning at all of the \texttt{HERWIG}'s parameters and with a reasonable choice of the \texttt{MCSTHAR++}'s parameters: $\gamma_{s}=0.65$ and \mbox{$\rho=0.35 \ \tx{GeV}/\tx{fm}^{3}$}. The distributions and the multiplicity table show a comparison among \texttt{MCSTHAR++}, \texttt{HERWIG6.510} and LEP (OPAL \cite{ref:opal} and DELPHI \cite{ref:delphi}) data for the process $e^{+}e^{-}\rightarrow \gamma/Z^{0}\rightarrow u\bar{u},d\bar{d},s\bar{s}$ at $91.2\ \tx{GeV}$ center of mass energy.

\begin{figure}
  \centering
  \subfigure{\includegraphics[width=0.45\textwidth]{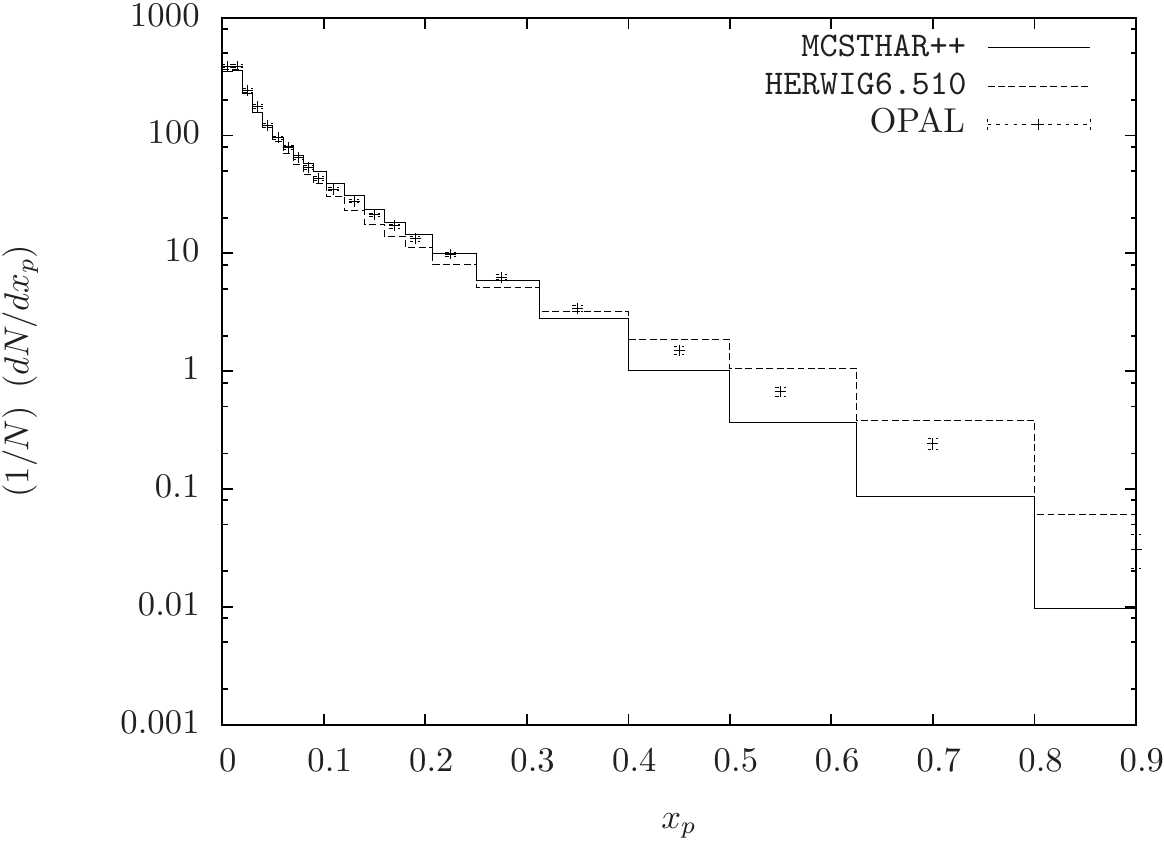}}                
  \subfigure{\includegraphics[width=0.45\textwidth]{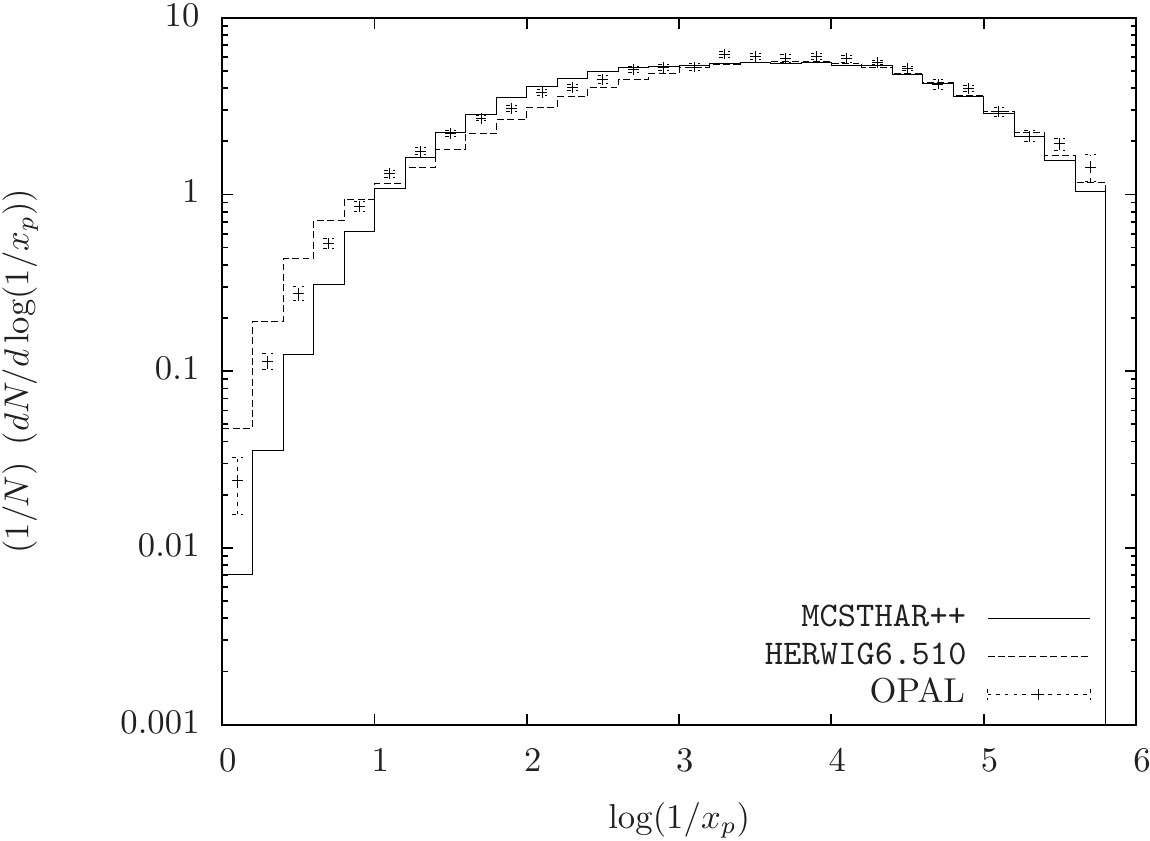}}
  \caption{Charged particle scaled momentum $x_{p} = 2p/\sqrt{s}$ and $log{\left(1/x_{p}\right)}$ distribution.
  % Comparison among
  %\texttt{HERWIG6.510} with \texttt{MCSTHAR++} as hadronization module, \texttt{HERWI6.510} and OPAL data \cite{ref:opal}.}
 }
  \label{fig:distr}
\end{figure}

\begin{scriptsize}
\begin{table}[!h]
 \caption{Mean values of the charged particle multiplicity and of the production rate of charged pions, charged kaons and (anti)protons, for the hadronization of light quarks only. 
% Comparison among %\texttt{HERWIG6.510} with \texttt{MCSTHAR++} as hadronization module, \texttt{HERWIG6.510} and DELPHI data \cite{ref:delphi}.}
}
  \label{tab:mult}
  \begin{tabular}{lcccc}
    \hline
      & $ N_{\tx{ch}} $ & $N_{\pi^{\pm}}$ & $N_{K^{\pm}}$ & $N_{p,\bar{p}}$ \\ 
    \hline
     \texttt{MCSTHAR++} & $ 19.53 \pm 0.14 $ & $ 16.64 \pm 0.11 $ & $ 1.65 \pm 0.04 $ & $ 0.98 \pm 0.07 $ \\ 
    \hline
     \texttt{HERWIG6.510} & $ 18.601 \pm 0.006 $ & $ 15.022 \pm 0.006 $ & $ 1.628 \pm 0.002 $ & $ 1.736 \pm 0.002 $ \\ 
    \hline
     DELPHI & $ 19.94 \pm 0.34 $ & $16.84 \pm 0.87 $ & $ 2.02 \pm 0.07 $ & $ 1.07 \pm 0.05 $ \\ 
    \hline
  \end{tabular}
     \end{table}
\end{scriptsize}

\vspace{-0.85cm}
\section{Conclusions}
\texttt{MCSTHAR++}, a new Monte Carlo code for the Statistical Microcanonical Hadronization, has been presented. This code, written in \texttt{C++}, will be soon available to be used with the Monte Carlo event generators alternatively to the standard hadronization modules, as it has been shown in this paper, where \texttt{MCSTHAR++} has been interfaced to \texttt{HERWIG6.510}.

The next steps of this work will be focused on the tuning of \texttt{MCSTHAR++}, completed with the hadronization of $c$ and $b$ quarks, on LEP and Tevatron data, and on the comparison with other available hadronization models.

\acknowledgments
One of us (CB) would like to thank S. Gieseke for support and useful discussions and the ITP of the Karslruhe University for warm hospitality.

\vspace{-0.45cm}

%\acknowledgments
%This work was produced, supported and perpetrated by S. Frabetti under
%the auspices of the Italian Physical Society.
%Grazie to M. Missiroli for the valuable collaboration.


\begin{thebibliography}{0}
\bibitem{ref:CluMod} 
	\BY{Webber,~B.R.} \IN{Nucl. Phys. B}{\textbf{238}}{1984}{492};
  	\BY{Marchesini,~G. \atque Webber,~B.R.} \IN{Nucl. Phys. B}{\textbf{310}}{1988}{461}. 

\bibitem{ref:SherpMod}
 	\BY{Winter,~J.C., Krauss,~F. \atque Soff,~G.} \IN{Eur. Phys. J. C}{\textbf{36}}{2004}{381}. 

\bibitem{ref:StrMod} 
  	\BY{Andersson,~B., Gustafson,~G., Ingelman,~G. \atque Sj\"ostrand,~T.} \IN{Phys. Rep.}{\textbf{97}}{1983}{31}. 

\bibitem{ref:HagBec}
%	\BY{Hagedorn,~R.} \IN{Nuovo Cim. Suppl.}{\textbf{3}}{1965}{147};
%  	\BY{Becattini,~F.} \IN{Z. Phys. C}{\textbf{69}}{1996}{485}; 
%	\BY{Becattini,~F. \atque Heinz,~U.W.} \IN{Z. Phys. C}{\textbf{76}}{1997}{269}
	\BY{Becattini,~F.} \IN{arXiv:0901.3643 [hep-ph]}{}{2009}{}, and references therein.

  
\bibitem{ref:Herwig65}
	\BY{Corcella,~G., Knowles,~I.G., Marchesini,~G., Moretti,~S., Odagiri,~K., Richardson,~P., Seymour,~M.H. \atque Webber,~B.R.} \IN{JHEP}{\textbf{0101}}{2001}{010}.

\bibitem{ref:BecFer}
	\BY{Becattini,~F. \atque Ferroni,~L.} \IN{Eur. Phys. J. C}{\textbf{35}}{2004}{243}; \SAME{38}{2004}{225}.

\bibitem{ref:opal}
	\BY{Ackerstaff,~K. \emph{et al.}} \IN{Eur. Phys. J. C}{\textbf{7}}{1999}{369}.
	
\bibitem{ref:delphi}
	\BY{Abreu,~P. \emph{et al.}} \IN{Eur. Phys. J. C}{\textbf{5}}{1998}{585}.

	
%\bibitem{ref:} \BY{Boccaccio~G. \atque de~Cam\~oes~L.}
%  \IN{Phys. Rev. A}{13}{1999}{12};
%  \SAME{69}{999}{1666}.
%\bibitem{ref:pul} \BY{Pulci~L.}
%  preprint INFN 8181.
%\bibitem{ref:bra} \BY{Bragg~B.}
%  \TITLE{Tender comrade},
%  in \TITLE{Workers Playtime},
%                  edited by \NAME{Tizio A. \atque Caio B.}
%                  (Unexeditor, Bologna) 1997, pp.~1-10.
\end{thebibliography}
\end{document}